\documentclass[11pt,reprint]{revtex4-2} 
\usepackage{graphicx} 
\usepackage{amsmath,bm}
\usepackage{hyperref}
\usepackage{xcolor}
\usepackage[capitalize]{cleveref}

\newcommand{\ld}{\lambda_{\mathrm{D}}}

\newcommand{\nn}{\nonumber\\}

\usepackage{cancel}

\begin{document}
\title{Impedance of an electric double layer capacitor with a multi-component electrolyte}

 \author{David Fertig}
 \email{david.fertig@nmbu.no}
 \affiliation{Institute of Physics, Norwegian University of Life Sciences, \AA s, Norway}
\date{\today}

\begin{abstract}
I derive the impedance response of an ideal electrolyte containing an arbitrary number of mobile ionic species between blocking planar electrodes, described by the Poisson--Nernst--Planck equations.
By transforming the linearized equations to a charge--salt basis, the response is written in terms of a multi-component diffusion--migration matrix and its eigenvalues/eigenvectors.
When all diffusivities are equal, the charge mode decouples from the neutral concentration subspace and the classical binary-electrolyte result is recovered.
In contrast, unequal diffusivities couple charge relaxation to one or more neutral composition modes.
For a ternary electrolyte with two cations and one anion, this coupling produces additional diffusive features and broadens the crossover between resistive and capacitive regimes.
The results found in this paper provide a minimal continuum explanation for why mixed electrolytes can display impedance spectra that cannot be interpreted as a simple binary electrolyte with an averaged diffusion coefficient.
\end{abstract}

\maketitle
\section{Introduction}
Electrochemical impedance spectroscopy is a useful tool for probing ion transport, interfacial polarization, and charge storage in electrolyte-filled cells~\cite{Orazem,Lasia}.
In the small-amplitude regime, the measured impedance reflects the linear response of ionic concentrations and electric potential to an applied oscillatory perturbation.
Continuum models, particularly the Poisson--Nernst--Planck (PNP) framework, provide theoretical basis for understanding the frequency-dependent impedance of the electrolytic cell~\cite{usler_arxiv_2026}.

Multi-component systems, although rarely investigated in model electric double-layer capacitors~\cite{horno_jec_1996,jarvey_sm_2023}, are important in widely different contexts.
Physiological solutions, besides containing a plethora of organic materials, contain sodium, potassium, magnesium, and chloride ions.
Seawater, which can be used for blue-energy harvesting~\cite{janssen_prl_2014}, also contains a wide variety of ions, although it is dominated by sodium chloride.
Systems with supporting electrolytes~\cite{dickinson_jpcc_2009} also contain at least three ionic species.
Similarly, redox-flow batteries, such as vanadium redox-flow batteries~\cite{heiss_ea_2024} often involve multiple ionic components, making multi-component electrolyte effects relevant well beyond idealized binary systems.

Continuum electrodiffusion theories based on PNP-type equations~\cite{Bazant2004,Kilic2007a,Kilic2007b,janssen_pre_2018}, and modified double-layer theories for concentrated electrolytes and ionic liquids~\cite{Kornyshev2007,Fedorov2014} have been used to investigate the equilibrium/steady-state and disentangle the transient responses over various timescales.
The diffuse-charge dynamics framework of Bazant, Thornton, and Ajdari is especially relevant, as it connects the frequency-dependent response of blocking or polarizable electrodes to bulk ion transport and capacitive charging of the electric double layer~\cite{Bazant2004}.
For a symmetric binary electrolyte, the linearized dynamics can be written in terms of a charge density and a single neutral salt density, which are decoupled, and an analytical expression for the electrostatic potential can be obtained~\cite{janssen_pre_2018}.
The apparent simplicity of binary electrolytes is partly a consequence of the small number of independent concentration variables.
Ternary electrolytes represent an important intermediate case between binary models and fully general multi-component transport, while still retaining some analytical tractability.
Horno et al.~\cite{horno_jec_1996} analyzed a three-component system with blocking electrodes and equal diffusion coefficients at different biases.
Ref.~\cite{jarvey_sm_2023} also considered a multi-component, although reactive system.

Electric double layer capacitors (EDLC) are especially common systems for impedance analysis because their charge-storage mechanism is dominated, to a first approximation, by reversible ion electrosorption and double-layer charging rather than by slow Faradaic reactions.
Their impedance has therefore been analyzed starting since the pioneering work of Macdonald~\cite{Macdonald1953} who studied the admittance (inverse of the impedance) of systems with mobile charge carriers blocked at the electrodes. 
Classical porous-electrode and transmission-line models, originating from de Levie's theory, describe the distributed resistance and capacitance associated with ion motion inside porous electrodes~\cite{DeLevie1963,Itagaki2007}.
Such models are widely used to interpret the characteristic transition from a high-frequency resistive response, through a Warburg-like or transmission-line region, to a low-frequency capacitive response in porous carbon EDLCs.
Barbero and co-workers provided another important perspective by emphasizing how surface adsorption--desorption processes and generalized Langmuir-type adsorption kinetics can modify the low-frequency impedance of electrolyte-filled cells and EDLC-like systems~\cite{Barbero2006,Barbero2007,AlexeIonescu2019}.
Related EDLC studies have included analyses of sub-nanometer carbon pores, hierarchical porous electrodes, and porous-electrode-theory predictions of EDLC charge--discharge and impedance behavior~\cite{Segalini2010,Hasyim2017,Allagui2022}.
Interestingly, the admittance response can also be sampled with molecular dynamics simulations~\cite{pireddu2024impedance} via transient electrode charge correlations.

Although linearized PNP models are well-developed for binary electrolytes, the corresponding impedance structure for ternary, quaternary, and more general multi-component systems is less commonly presented in explicit form.
The purpose of the present work is not to introduce new interfacial physics, but to isolate the consequences of multi-component transport within the simplest possible continuum model.
I therefore consider an ideal electrolyte between blocking planar electrodes.
This excludes finite ion-size effects, ion--ion correlations, Stern layers, and Faradaic reactions.
Within this minimal setting, any deviation from the classical impedance of binary electrolytes can be attributed directly to the presence of additional mobile species and to differences in their diffusion coefficients.

The paper is structured as follows. In \cref{sec:model}, I introduce the setup and the governing equations and show how impedance is calculated.
\Cref{sec:general} details the derivation of the impedance response for an $N$-component electrolyte, whereas in \cref{sec:tern}, I derive the impedance response for a ternary system.
In \cref{sec:discussion}, I discuss the results for the ternary system, and I conclude the article in \cref{sec:concl}.

\section{Model}\label{sec:model}
\subsection{Setup}
I consider an electrolyte with $N$ different ionic components, where $z_i$ are the valencies (\mbox{$i=1,\dots,N$}), $D_i$ are the diffusion coefficients, and $X_i$ are the stoichiometric coefficients of the cations and anions ($\sum_i z_iX_i=0$).
The bulk concentration of the individual species is $X_ic_0$.
The ionic strength of the electrolyte is $I=\frac{1}{2}\sum_i X_iz_i^2c_0$, the temperature is $T$, and the electrolyte is between two parallel flat, blocking electrodes separated by a distance $2L$. The Cartesian coordinate $x$ runs from the left ($x=-L$) to the right ($x=L$). I apply a potential difference $2\Psi$ at $t=0$,
\begin{align}\label{bc:pot}
    \psi\big|_{x=\pm L,t}=\pm \Psi.
\end{align}
\subsection{Governing equations}\label{sec:gov}
The Poisson equation is
\begin{align}
    \partial_x^2\psi=-\dfrac{eq}{\varepsilon_0\varepsilon_r},
\end{align}
where $q=\sum_iz_i\rho_i$ is the charge density, and $e$ is the elementary charge.
The ionic fluxes are described by the Nernst-Planck equation,
\begin{align}
j_i &= -D_i\left(\partial_x\rho_i +\dfrac{z_i e}{kT}\rho_i\partial_x\psi\right).
\end{align}
The applied potential is less than the thermal voltage $\tilde{\Psi}=e\Psi/kT\ll 1$, which allows the linearization of the problem using an asymptotic expansion. 
I write the individual densities and the potential as
$\rho_i=X_ic_0+\tilde{\Psi}\rho_i^{(1)}+\mathcal{O}(\tilde{\Psi}^2)$, $\psi=0+\tilde{\Psi}\psi^{(1)}+\mathcal{O}(\tilde{\Psi}^2)$. At $\mathcal{O}(\tilde{\Psi})$, I find
\begin{align}
j^{(1)}_i &= -D_i\left(\partial_x\rho^{(1)}_i +\dfrac{z_i e X_i}{kT}c_0\partial_x\psi^{(1)}\right).
\end{align}
For better readability, I omit the superscript $(1)$ from now on.
The continuity equation then reads
\begin{align}
    \partial_t\rho_i &= -\partial_x j_i = D_i\left(\partial^2_x\rho_i +\dfrac{z_ie X_i}{kT}c_0\partial^2_x\psi\right).
\end{align}
Using $\tilde{x}=x/\lambda_{D}$, $\tilde{\rho}=\rho/\Gamma$ with $\Gamma=c_0\sum\limits_i z_i^2X_i$, and $\tilde{\psi}=e\psi/kT$, and $\tilde{t}=tD_1/\lambda_D^2$, the nondimensional formulation of the continuity equation reads
\begin{align}\label{eq:continuity}
    \partial_t\tilde{\rho}_i &=  \epsilon_i\left(\partial^2_{\tilde{x}}\tilde{\rho}_i +\dfrac{z_iX_i}{S}\partial^2_{\tilde{x}}\tilde{\psi}\right),
\end{align}    
with $\epsilon_i=D_i/D_1$, $\epsilon_1=1$.
The Poisson equation in nondimensional form is then
\begin{align}\label{eq:nondimpoiss}
    \partial_{\tilde{x}}^2\tilde{\psi}=-\tilde{q}.
\end{align}
where $\tilde{q}=q/\Gamma$ is the dimensionless charge density.
The equations are solved with the following boundary conditions:
At time $\tilde{t}=0$, $\tilde{\rho}_i(\tilde{x},\tilde{t})=X_ic_0/\Gamma$.
I apply a potential step at $\tilde{x}=\pm\tilde{L}$, with $\tilde{L}=L/\ld$ at $\tilde{t}=0$, $\tilde{\psi}(\pm\tilde{L},0)=\pm\tilde{\Psi}$. The electrodes are blocking, i.e. $\tilde{j}_i(\pm\tilde{L},\tilde{t})=0$ for all ionic species.
\subsection{Impedance}
The impedance is defined as
\begin{align}
    Z=\dfrac{\hat{\tilde{\psi}}(\tilde{x}=\tilde{L})}{\hat{\tilde{\iota}}(\tilde{x}=\tilde{L})}
\end{align}
where the hat denotes the Laplace transform $\hat{f}(\tilde{p})=\int_0^{\infty}e^{-\tilde{p}\tilde{t}}f(\tilde{t})\,\mathrm{d}\tilde{t}$, $\tilde{p}=p\ld^2/D_1$ is the dimensionless Laplace variable, and $\tilde{\iota}$ is the dimensionless areal electronic current defined as $\tilde{\iota}=\partial_{\tilde{t}}\partial_{\tilde{x}}\tilde{\psi}$.
Substituting the boundary condition \cref{bc:pot} and carrying out the Laplace transformation, one finds at $\tilde{x}=\tilde{L}$
\begin{align}\label{eq:impedance}
    Z=\dfrac{\tilde{\Psi}/\tilde{p}}{\tilde{p}\partial_{\tilde{x}}\hat{\tilde{\psi}}(\tilde{x}=\tilde{L})-\cancel{\tilde{\iota}(\tilde{t}=0)}}.
\end{align}
\section{General solution of the PNP equations for an $N$-component system}\label{sec:general}
Instead of solving \cref{eq:continuity,eq:nondimpoiss} with the individual ionic densities, I switch to a charge-salt description.
Contrary to binary electrolytes where the charge-salt picture is simple~\cite{janssen_pre_2018}, for $N$ different components, one has multiple ways of defining the salt basis.
The choice is not entirely arbitrary, the elements of the new salt base have to be linearly independent as well as electroneutral.
I build the new basis as 
\begin{align}\label{eq:saltcharge}
    \begin{pmatrix}
\tilde{q} \\
\tilde{s}_1\\
\tilde{s}_2\\
\vdots\\
\tilde{s}_{N-1}\\
\end{pmatrix}=
\begin{pmatrix}
z_1 & z_2 & \dots & z_N \\
\nu_{11} & \nu_{12} & \dots & \nu_{1N}\\
\nu_{21} & \nu_{22} & \dots & \nu_{2N}\\
\vdots &  \ddots & & \vdots\\
\nu_{N-1,1} & \nu_{N-1,2} & \dots & \nu_{N-1,N}
\end{pmatrix}
\begin{pmatrix}
\tilde{\rho}_1 \\
\tilde{\rho}_2\\
\tilde{\rho}_3\\
\vdots\\
\tilde{\rho}_N\\
\end{pmatrix},
\end{align}
where $\tilde{s}_a$, $a=1,\dots,N-1$ are the dimensionless salt variables.
The coefficients $\nu_{ai}$ define $N-1$ linearly independent electroneutral
linear combinations of the density perturbations. In particular, the row
vectors $\boldsymbol{\nu}_{a+1}=(\nu_{a1},\nu_{a2},\dots,\nu_{aN})$
are chosen to span the subspace orthogonal to the charge vector
$\mathbf{z}=(z_1,z_2,\dots,z_N)$, i.e.
\begin{align}
\sum_{i=1}^N z_i \nu_{ai}=0,
\quad a=1,\dots,N-1.    
\end{align}
I introduce $\mathbf{w}=(\tilde{q},\tilde{s}_1,\dots,\tilde{s}_{N-1})^T$ where $T$ denotes the transpose operation, $\mathbf{C}=(\mathbf{z},\boldsymbol{\nu}_1,\dots,\boldsymbol{\nu}_{N-1})^T$, $\boldsymbol{\rho}=(\tilde{\rho}_1,\dots,\tilde{\rho}_N)^T$, allowing to rewrite \cref{eq:saltcharge} as $\mathbf{w}=\mathbf{C}\boldsymbol{\rho}$.
Before continuing, I rewrite \cref{eq:continuity} as
\begin{align}
\partial_{\tilde{t}}\boldsymbol{\rho}=\boldsymbol{\epsilon}\partial_{\tilde{x}}^2\boldsymbol{\rho}+\boldsymbol{\epsilon}\mathbf{f}\partial_{\tilde{x}}^2\tilde{\psi}
\end{align}
with $\boldsymbol{\epsilon}=\mathrm{diag}(\epsilon_1,\dots,\epsilon_N)$ and $\mathbf{f}=(z_1X_1,\dots,z_NX_N)^T/S$.
I now write $\boldsymbol{\rho}$ as $\mathbf{C}^{-1}\mathbf{w}$.
Multiplying both sides with $\mathbf{C}$ results in
\begin{align}
\partial_{\tilde{t}}\mathbf{w}=\mathbf{C}\boldsymbol{\epsilon}\mathbf{C}^{-1}\partial_{\tilde{x}}^2\mathbf{w}+\mathbf{C}\boldsymbol{\epsilon}\mathbf{f}\partial_{\tilde{x}}^2\tilde{\psi}.
\end{align}
Introducing $\mathbf{M}=\mathbf{C}\boldsymbol{\epsilon}\mathbf{C}^{-1}$, $\mathbf{L}=\mathbf{C}\boldsymbol{\epsilon}\mathbf{f}$ and using \cref{eq:nondimpoiss}
\begin{align}
    \partial_{\tilde{t}}\mathbf{w}=\mathbf{M}\partial_{\tilde{x}}^2\mathbf{w}-\mathbf{L}\tilde{q}.
\end{align}
As $\tilde{q}=\mathbf{e}_{\tilde{q}}^T\mathbf{w}$ with $\mathbf{e}_{\tilde{q}}=(1,0,\dots,0)^T$, I write $\mathbf{B}=\mathbf{L}\mathbf{e}_{\tilde{q}}^T$
\begin{align}\label{eq:matrixcont}
    \partial_{\tilde{t}}\mathbf{w}=\mathbf{M}\partial_{\tilde{x}}^2\mathbf{w}-\mathbf{B}\mathbf{w}.
\end{align}
I Laplace transform \cref{eq:matrixcont}
\begin{align}
    \tilde{p}\hat{\mathbf{w}}-\mathbf{w}_0=\mathbf{M}\partial_{\tilde{x}}^2\hat{\mathbf{w}}-\mathbf{B}\hat{\mathbf{w}}, 
\end{align}
with $\mathbf{w}_0=(0,\sum_a\nu_{1i}X_i,\sum_i\nu_{2i}X_i,\dots,\sum_i\nu_{N-1,i}X_i)c_0/\Gamma$. Introducing the modified salt variables, $\hat{\tilde{u}}_a=\hat{\tilde{s}}_a-\frac{1}{\tilde{p}}\frac{\sum_i \nu_{ai}X_ic_0}{\Gamma}$ and using $\hat{\mathbf{v}}=(\hat{\tilde{q}},\hat{\tilde{u}}_1,\dots,\hat{\tilde{u}}_{N-1})^T$, I find
\begin{align}\label{eq:eigen}
\partial_{\tilde{x}}^2\hat{\mathbf{v}}=\boldsymbol{\Lambda}\hat{\mathbf{v}}, 
\end{align}
where $\boldsymbol{\Lambda}=\mathbf{M}^{-1}(\tilde{p}\mathbf{I}+\mathbf{B})$ and $\mathbf{I}$ is the identity matrix.
Diagonalization of $\boldsymbol{\Lambda}$ lets one find its eigenvalues and with these eigenvalues and eigenvectors, the system of differential equations decouples.
I obtain $N$ ordinary differential equations
\begin{align}\label{eq:diag}
    \partial_{\tilde{x}}^2\hat{\tilde{y}}_i=\lambda_i \hat{\tilde{y}}_i
\end{align}
where $\lambda_i$ are the eigenvalues of $\boldsymbol{\Lambda}$ for $i=1,\dots,N$.
Introducing $k_{i}=\sqrt{\lambda_{i}}$, I find
\begin{subequations}
\begin{align}
    \hat{\tilde{y}}_i &= A_{i}\sinh(k_i \tilde{x})+B_{i}\cosh(k_i \tilde{x}).
\end{align}    
\end{subequations}

I reconstruct $\mathbf{\hat{v}}$ as $\sum_i \hat{\tilde{y}}_i\boldsymbol{\alpha}_i$, where $\boldsymbol{\alpha}_i$ are the eigenvectors of $\boldsymbol{\Lambda}$
\begin{align}
    \mathbf{\hat{v}}&=\sum\limits_i[A_{i}\sinh(k_i \tilde{x})+B_{i}\cosh(k_i \tilde{x})]\boldsymbol{\alpha}_i.
\end{align}
The charge and salt modes are then
\begin{subequations} \label{eq:q_s}   
\begin{align}
    \hat{\tilde{q}}&=\sum\limits_i[A_{i}\sinh(k_i \tilde{x})+B_{i}\cosh(k_i \tilde{x})]\alpha_{i1},\\
    \hat{\tilde{s}}_l&=\dfrac{\sum_a\nu_{l+1,a}X_ac_0}{\tilde{p}\Gamma}+\sum\limits_i[A_{i}\sinh(k_i \tilde{x})+B_{i}\cosh(k_i \tilde{x})]\alpha_{i,l+1},
\end{align}
\end{subequations}
where $\alpha_{ij}$ is the $j$-th element of $\boldsymbol{\alpha}_i$, $i=1,\dots,N$, $l=1,\dots,N-1$.
I integrate the charge profiles twice to obtain the electrostatic potential profile
\begin{align}\label{eq:pot}
    \hat{\tilde{\psi}}=-\sum\limits_i\left[\dfrac{A_{i}}{k_i^2}\sinh(k_i \tilde{x})+\dfrac{B_{i}}{k_i^2}\cosh(k_i \tilde{x})\right]\alpha_{i1}+C_1\tilde{x}+C_0.
\end{align}
The Laplace transformed boundary conditions for the charge-salt picture are
$\hat{\tilde{\psi}}=\pm\tilde{\Psi}/\tilde{p}$, $\partial_{\tilde{x}}\hat{\tilde{q}}+\partial_{\tilde{x}}\hat{\tilde{\psi}}=0$, and $\partial_{\tilde{x}}\hat{\tilde{s}}=0$.
These boundary conditions follow by applying the same charge--salt transformation to the individual blocking conditions $\tilde{j}_i=0$.
Considering the symmetry of the system, I find $B_i=0$ and $C_0=0$.
Combining the boundary conditions with \cref{eq:q_s,eq:pot} result in the equations
\begin{subequations}\label{eq:bceqs}
    \begin{align}
        &\sum\limits_i A_i k_i\cosh(k_i\tilde{L})\alpha_{i,j+1}=0\quad j=1,\dots,N-1,\\
        &C_1+\sum\limits_i A_i\left(k_i-\dfrac{1}{k_i}\right)\cosh(k_i\tilde{L})\alpha_{i,1}=0,\\
        &C_1\tilde{L}-\sum\limits_i\dfrac{A_i}{k_i^2}\sinh(k_i\tilde{L})\alpha_{i,1}=\dfrac{\tilde{\Psi}}{\tilde{p}}.
    \end{align}
\end{subequations}
In matrix form
    \begin{align}
        \begin{pmatrix}
\boldsymbol{\alpha}_s\mathbf{S}_1 & \mathbf{0} \\
\boldsymbol{\alpha}_q\mathbf{S}_2  & 1 \\
-\boldsymbol{\alpha}_q\mathbf{S}_3  & \tilde{L}
\end{pmatrix}
\begin{pmatrix}
    \mathbf{A}\\
    C_1
\end{pmatrix}=
\begin{pmatrix}
    \mathbf{0}\\
    0\\
    \tilde{\Psi}/\tilde{p}
\end{pmatrix},
    \end{align}
with 
\begin{subequations}
    \begin{align}
        \mathbf{S}_1&=\mathrm{diag}\bm(k_1\cosh(k_1\tilde{L}),\dots,k_N\cosh(k_N\tilde{L})\bm),\\
        \mathbf{S}_2&=\mathrm{diag}\bm(\left(k_1-\dfrac{1}{k_1}\right)\cosh(k_1\tilde{L}),\dots, \left(k_N-\dfrac{1}{k_N}\right)\cosh(k_N\tilde{L})\bm),\\
        \mathbf{S}_3&=\mathrm{diag}\bm(\dfrac{\sinh(k_1\tilde{L})}{k_1^2},\dots,\dfrac{\sinh(k_N\tilde{L})}{k_N^2}\bm),\\
        \mathbf{A}&=(A_1,\dots,A_N)^T,\\
        \boldsymbol{\alpha}_q&=(\alpha_{11},\dots,\alpha_{N1}),\\
        \boldsymbol{\alpha}_s&=\begin{pmatrix}
\alpha_{12} & \alpha_{22} & \dots & \alpha_{N2} \\
\alpha_{13} & \alpha_{23} &  \dots & \alpha_{N3}\\
\vdots & & \ddots & \vdots\\
\alpha_{1N} & \alpha_{2N} & \dots & \alpha_{NN}
\end{pmatrix}.
    \end{align}
\end{subequations}  
The matrix equation $\boldsymbol{\alpha}_s\mathbf{S}_1\mathbf{A}=0$ has $N-1$ equations and $N$ unknown parameters.
One can fix one of them, e.g. $A_1$ and express the rest as a function of the chosen parameter.
Then $\mathbf{A}$ can be expressed as $\mathbf{A}=A_1\boldsymbol{\eta}$, where $\boldsymbol{\eta}$ is
\begin{align}\label{eq:matrix_eta}
    \boldsymbol{\eta}=\begin{pmatrix}
        1\\
        -\dfrac{k_1\cosh(k_1\tilde{L})}{k_2\cosh(k_2\tilde{L})}\dfrac{\det(\boldsymbol{\alpha}_s^{(2)})}{\det(\boldsymbol{\alpha}_s^{(1)})}\\
        \vdots\\
        -\dfrac{k_1\cosh(k_1\tilde{L})}{k_N\cosh(k_N\tilde{L})}\dfrac{(-1)^{N}\det(\boldsymbol{\alpha}_s^{(N)})}{\det(\boldsymbol{\alpha}_s^{(1)})}
    \end{pmatrix},
\end{align}
where the index $(i)$ in the superscript denotes which column is left out from $\boldsymbol{\alpha}_s$.
In \cref{appendix:eta}, I show the construction of $\boldsymbol{\eta}$ for simpler systems.

$C_1$ is then expressed as
\begin{align}\label{eq:c1}
    C_1=-A_1\boldsymbol{\alpha_q}\mathbf{S}_2\boldsymbol{\eta}.
\end{align}
Then I find $A_1$ with
\begin{align}
    -\boldsymbol{\alpha}_q\mathbf{S}_3\mathbf{A}+C_1\tilde{L}=\dfrac{\tilde{\Psi}}{\tilde{p}},
\end{align}
which I combine with \cref{eq:c1} to find
\begin{align}\label{eq:a1}
    A_1=-\dfrac{\tilde{\Psi}}{\tilde{p}}\dfrac{1}{\boldsymbol{\alpha}_q\mathbf{S}_3\boldsymbol{\eta}+\boldsymbol{\alpha}_q\mathbf{S}_2\boldsymbol{\eta}\tilde{L}}
\end{align}
and obtain
\begin{align}\label{eq:a}
    \mathbf{A}=A_1\boldsymbol{\eta}=-\dfrac{\tilde{\Psi}}{\tilde{p}}\dfrac{\boldsymbol{\eta}}{\boldsymbol{\alpha}_q\mathbf{S}_3\boldsymbol{\eta}+\boldsymbol{\alpha}_q\mathbf{S}_2\boldsymbol{\eta}\tilde{L}}.
\end{align}
I construct the Laplace-transformed potential using \cref{eq:c1,eq:a},
\begin{align}\label{eq:potential}
    \hat{\tilde{\psi}}(\tilde{x},\tilde{p})=\dfrac{\tilde{\Psi}}{\tilde{p}}\dfrac{\boldsymbol{\alpha}_q\mathbf{S}_x\boldsymbol{\eta}+\boldsymbol{\alpha_q}\mathbf{S}_2\boldsymbol{\eta}\tilde{x}}{\boldsymbol{\alpha}_q\mathbf{S}_3\boldsymbol{\eta}+\boldsymbol{\alpha}_q\mathbf{S}_2\boldsymbol{\eta}\tilde{L}},
\end{align}
with $\mathbf{S}_x=\mathrm{diag}\Big(\dfrac{\sinh(k_1\tilde{x})}{k_1^2},\dots,\dfrac{\sinh(k_N\tilde{x})}{k_N^2}\Big)$.
The spatial derivative of \cref{eq:potential} gives,
\begin{align}
    \partial_{\tilde{x}}\hat{\tilde{\psi}}|_{\tilde{x}=\tilde{L}}=\dfrac{\tilde{\Psi}}{p}\dfrac{\boldsymbol{\alpha}_q\mathbf{S}_1\boldsymbol{\eta}}{\boldsymbol{\alpha}_q\mathbf{S}_3\boldsymbol{\eta}+\boldsymbol{\alpha}_q\mathbf{S}_2\boldsymbol{\eta}\tilde{L}},
\end{align}
from which I obtain the impedance of the system using \cref{eq:impedance}
\begin{align}\label{eq:general_impedance}
Z=\dfrac{\boldsymbol{\alpha}_q\mathbf{S}_3\boldsymbol{\eta}+\boldsymbol{\alpha}_q\mathbf{S}_2\boldsymbol{\eta}\tilde{L}}{\tilde{p}\boldsymbol{\alpha}_q\mathbf{S}_1\boldsymbol{\eta}}.
\end{align}
Although \cref{eq:general_impedance} gives the impedance of any electrolyte system subject to a linear perturbation, due to the nature of the vectors and matrices (such as $\boldsymbol{\eta}$, as its elements require to calculate the determinant of an $N-1\times N-1$ matrix), it is quite difficult to analytically analyze the expression due to the large amount of terms (numerical investigations are of course possible).
Therefore, in the next section I re-examine the equations for a simpler system, a ternary electrolyte.
\section{Impedance of a ternary electrolyte}\label{sec:tern}
For the sake of simplicity, I use a simple, equimolar mixture of salts, with monovalent components: $z_1=1$, $z_2=1$, $z_3=-1$, $X_{1}=1$, $X_{2}=1$, $X_{3}=2$. \Cref{appendix:general} continues with a general description of the equations. \Cref{eq:continuity} then simplifies to
\begin{subequations}
\begin{align}
    \partial_{\tilde{t}}\tilde{\rho}_1 &=  \partial^2_{\tilde{x}}\tilde{\rho}_1 +\dfrac{1}{4}\partial^2_{\tilde{x}}\tilde{\psi}\\
    \partial_{\tilde{t}}\tilde{\rho}_2 &=  \epsilon_2\left(\partial^2_{\tilde{x}}\tilde{\rho}_2 +\dfrac{1}{4}\partial^2_{\tilde{x}}\tilde{\psi}\right)\\
    \partial_{\tilde{t}}\tilde{\rho}_3 &=  \epsilon_3\left(\partial^2_{\tilde{x}}\tilde{\rho}_3 -\dfrac{1}{2}\partial^2_{\tilde{x}}\tilde{\psi}\right).
\end{align}
\end{subequations}
I introduce the salt variables $s_1=2\rho_1+\rho_3$ and $s_2=2\rho_2+\rho_3$ and then express the individual ionic densities using the new charge-salt density basis
\begin{subequations}
\begin{align}
    \tilde{\rho}_1 &= \dfrac{1}{4}\tilde{q}+\dfrac{3}{8}\tilde{s}_1-\dfrac{1}{8}\tilde{s}_2,\\
    \tilde{\rho}_2 &= \dfrac{1}{4}\tilde{q}-\dfrac{1}{8}\tilde{s}_1+\dfrac{3}{8}\tilde{s}_2,\\
    \tilde{\rho}_3 &= -\dfrac{1}{2}\tilde{q}+\dfrac{1}{4}\tilde{s}_1+\dfrac{1}{4}\tilde{s}_2.
\end{align}    
\end{subequations}
I rewrite the continuity equations with using charge--salt basis
\begin{subequations}\label{eq:final}
\begin{align}
    \partial_{\tilde{t}} \tilde{q} &=  \dfrac{1+\epsilon_2+2\epsilon_3}{4}\partial_{\tilde{x}}^2\tilde{q}+\dfrac{3-\epsilon_2-2\epsilon_3}{8}\partial_{\tilde{x}}^2\tilde{s}_1\nn
    &+\dfrac{-1+3\epsilon_2-2\epsilon_3}{8}\partial_{\tilde{x}}^2\tilde{s}_2 +\dfrac{1+\epsilon_2+2\epsilon_3}{4}\partial^2_{\tilde{x}}\tilde{\psi}\\
    \partial_{\tilde{t}} \tilde{s}_1 &= \dfrac{1-\epsilon_3}{2}\partial_{\tilde{x}}^2\tilde{q}+\dfrac{3+\epsilon_3}{4}\partial_{x}^2\tilde{s}_1+\dfrac{\epsilon_3-1}{4}\partial_{\tilde{x}}^2\tilde{s}_2 \nn
    &+\dfrac{1-\epsilon_3}{2}\partial^2_{\tilde{x}}\tilde{\psi}\\
    \partial_{\tilde{t}} \tilde{s}_2 &= \dfrac{\epsilon_2-\epsilon_3}{2}\partial_{\tilde{x}}^2\tilde{q}+\dfrac{\epsilon_3-\epsilon_2}{4}\partial_{\tilde{x}}^2\tilde{s}_1+\dfrac{3\epsilon_2+\epsilon_3}{4}\partial_{\tilde{x}}^2\tilde{s}_2 \nn
    & +\dfrac{\epsilon_2-\epsilon_3}{2}\partial^2_{\tilde{x}}\tilde{\psi}.
\end{align}
\end{subequations}
First, I Laplace transform \cref{eq:final}, then insert \cref{eq:nondimpoiss}, and the use $\hat{\tilde{u}}_1=\hat{\tilde{s}}_1-\frac{1}{\tilde{p}}$ and $\hat{\tilde{u}}_2=\hat{\tilde{s}}_2-\frac{1}{\tilde{p}}$ results in
\begin{subequations}\label{eq:laplfinm1}
\begin{align}
    \tilde{p} \hat{\tilde{q}} &=  
    \dfrac{1+\epsilon_2+2\epsilon_3}{4}\partial_{\tilde{x}}^2\hat{\tilde{q}}
    +\dfrac{3-\epsilon_2-2\epsilon_3}{8}\partial_{\tilde{x}}^2\hat{\tilde{u}}_1\nn
    &+\dfrac{-1+3\epsilon_2-2\epsilon_3}{8}\partial_{\tilde{x}}^2\hat{\tilde{u}}_2 
    -\dfrac{\epsilon_1+\epsilon_2+2\epsilon_3}{4}\hat{\tilde{q}},\\
    \tilde{p} \hat{\tilde{u}}_1 &= 
    \dfrac{1-\epsilon_3}{2}\partial_{\tilde{x}}^2\hat{\tilde{q}}
    +\dfrac{3+\epsilon_3}{4}\partial_{\tilde{x}}^2\hat{\tilde{u}}_1+\dfrac{\epsilon_3-1}{4}\partial_{\tilde{x}}^2\hat{\tilde{u}}_2\nn
    &-\dfrac{1-\epsilon_3}{2}\hat{\tilde{q}},\\
    \tilde{p} \hat{\tilde{u}}_2 &= 
    \dfrac{\epsilon_2-\epsilon_3}{2}\partial_{\tilde{x}}^2\hat{\tilde{q}}
    +\dfrac{\epsilon_3-\epsilon_2}{4}\partial_{\tilde{x}}^2\hat{\tilde{u}}_1+\dfrac{3\epsilon_2+\epsilon_3}{4}\partial_{\tilde{x}}^2\hat{\tilde{u}}_2\nn
    &-\dfrac{\epsilon_2-\epsilon_3}{2}\hat{\tilde{q}}.
\end{align}
\end{subequations}

I rewrite \cref{eq:laplfinm1} in a matrix form with \mbox{$\hat{\mathbf{v}}=(\hat{\tilde{q}},\hat{\tilde{u}}_1,\hat{\tilde{u}}_2)^{T}$}
\begin{align}
    (\tilde{p}\mathbf{I}+\mathbf{B})\hat{\mathbf{v}} =\mathbf{M}\partial_{\tilde{x}}^2\hat{\mathbf{v}},
\end{align}
with
\begin{align}
    \mathbf{B}=
    \begin{pmatrix}
\dfrac{1+\epsilon_2+2\epsilon_3}{4} & 0 & 0 \\
\dfrac{1-\epsilon_3}{2} & 0 & 0 \\
\dfrac{\epsilon_2-\epsilon_3}{2} & 0 & 0
\end{pmatrix},
\end{align}
and
\begin{align}
    \mathbf{M}=
    \begin{pmatrix}
\dfrac{1+\epsilon_2+2\epsilon_3}{4} 
& \dfrac{3-\epsilon_2-2\epsilon_3}{8} 
& \dfrac{-1+3\epsilon_2-2\epsilon_3}{8} \\
\dfrac{1-\epsilon_3}{2} 
& \dfrac{3+\epsilon_3}{4} 
& \dfrac{\epsilon_3-1}{4} \\
\dfrac{\epsilon_2-\epsilon_3}{2} 
& \dfrac{\epsilon_3-\epsilon_2}{4} 
& \dfrac{3\epsilon_2+\epsilon_3}{4}
\end{pmatrix},
\end{align}or equivalently
\begin{align}\label{eq:lambda}
    \partial_{\tilde{x}}^2\mathbf{w}=\boldsymbol{\Lambda}\mathbf{w},
\end{align}
with $\boldsymbol{\Lambda}=\mathbf{M}^{-1}(\tilde{p}\mathbf{I}+\mathbf{B})$. 
The eigenvalues and the eigenvectors could be expressed analytically.
The characteristic polynomial of $\boldsymbol{\Lambda}$ has three solutions, which can be evaluated with the Cardano formula, although the analytical result does not provide sufficient insight.
Therefore I resort to solving the eigenvalue problem numerically. 
After diagonalizing matrix $\boldsymbol{\Lambda}$, and following \cref{eq:diag} to \cref{eq:potential}, I obtain 
\begin{widetext}
\begin{align}\label{eq:sol_phi}
   \hat{\tilde{\psi}}=\dfrac{\tilde{\Psi}}{\tilde{p}}\dfrac{\sinh(k_1\tilde{x})+\cosh(k_1 \tilde{L})\left(-\dfrac{f_{12}^3\sinh(k_2\tilde{x})}{\cosh(k_2 \tilde{L})}\Omega_1-\dfrac{f_{13}^3\sinh(k_3\tilde{x})}{\cosh(k_3 \tilde{L})}\Omega_2+\Upsilon k_1\tilde{x}\right)}{\sinh(k_1 \tilde{L})+\cosh(k_1 \tilde{L})\left(-f_{12}^3\tanh(k_2 \tilde{L})\Omega_1-f_{13}^3\tanh(k_3 \tilde{L})\Omega_2+\Upsilon k_1\tilde{L}\right)},
\end{align}
\end{widetext}
with the abbreviations \mbox{$\Omega_1=\dfrac{\alpha_{12}\alpha_{33}-\alpha_{13}\alpha_{32}}{\alpha_{22}\alpha_{33}-\alpha_{23}\alpha_{32}}\dfrac{\alpha_{21}}{\alpha_{11}}$}, \mbox{$\Omega_2=\dfrac{\alpha_{13}\alpha_{22}-\alpha_{23}\alpha_{12}}{\alpha_{22}\alpha_{33}-\alpha_{23}\alpha_{32}}\dfrac{\alpha_{31}}{\alpha_{11}}$}, $\Upsilon=k_1^2-1-\Omega_1\left(k_1^2-f_{12}^2\right)-\Omega_2\left(k_1^2-f_{13}^2\right)$, \mbox{$f_{12}=k_1/k_2$}, and \mbox{$f_{13}=k_1/k_3$}.
Note that parameters $\Omega_1$ and $\Omega_2$ are related to the elements of $\boldsymbol{\eta}$.
The impedance normalized by the bulk resistance $R_{\mathrm{bulk}}$ follows from \cref{eq:sol_phi}
\begin{align}\label{eq:sol_imp}
    \dfrac{Z}{R_{\mathrm{bulk}}} &=\bigg(\dfrac{\tanh(k_1 \tilde{L})-f_{12}^3\tanh(k_2 \tilde{L})\Omega_1-f_{13}^3\tanh(k_3 \tilde{L})\Omega_2}{\tilde{p}k_1^3(1-\Omega_1-\Omega_2)}\nn
    &+\dfrac{\Upsilon \tilde{L}}{\tilde{p}k_1^2(1-\Omega_1-\Omega_2)}\bigg)\dfrac{1}{R_{\mathrm{bulk}}},
\end{align}
with $R_{\mathrm{bulk}}=\tilde{L}/(1/4+1/4\epsilon_2+1/2\epsilon_3)$.
\Cref{eq:sol_imp} provides a general formula which can be simplified in the various limits.
For a model system with equal diffusivities ($\epsilon_i=1$), I find $\Omega_1=0$, $\Omega_2=0$ $R_{\mathrm{bulk}}=\tilde{L}$, and the expression in \cref{eq:sol_imp} simplifies to
\begin{align}\label{eq:Macdonald}
    \dfrac{Z}{R_{\mathrm{bulk}}} =\dfrac{\tanh(k_1 \tilde{L})+\left(k_1^2-1\right)k_1\tilde{L}}{\tilde{p}\tilde{L}k_1^3}=\dfrac{\tanh(k_1\tilde{L})}{\tilde{p}\tilde{L}k_1^3}+\dfrac{1}{k_1^2}
\end{align}
with $k_1=\sqrt{1+\tilde{p}}$.
\Cref{eq:Macdonald} is the impedance expression found by Macdonald~\cite{Macdonald1953} for a binary electrolyte with equal diffusion coefficients.
I would like to point out that for equal diffusivities, writing the equations in a matrix form is unnecessary, as the system of differential equations is decoupled from the start.

If, for example $\epsilon_2=\epsilon_3\neq 1$, one finds $\Omega_2=0$ and
\begin{align}
    \dfrac{Z}{R_{\mathrm{bulk}}} &=\bigg(\dfrac{\tanh(k_1 \tilde{L})-f_{12}^3\tanh(k_2 \tilde{L})\Omega_1}{\tilde{p}k_1^3\left(1-\Omega_1\right)}\nn
    &+\dfrac{[\left(k_1^2-1\right)-\Omega_1\left(k_1^2-f_{12}^2\right)]\tilde{L}}{\tilde{p}k_1^2\left(1-\Omega_1\right)}\bigg)\dfrac{1}{R_{\mathrm{bulk}}}.
\end{align}
A similar expression for binary electrolytes was found by Macdonald~\cite{Macdonald1953} and Usler et al.~\cite{usler_arxiv_2026}, and can be derived from the solution of Balu and Khair~\cite{balu2018role}.

\section{Discussion}\label{sec:discussion}
Most analytical treatments of linearized PNP impedance focus on these binary systems \cite{Barbero2007,Barbero2017, balu2018role,aslyamov2022analytical,Pedersen2023}, often with symmetric valences.
In a multi-component electrolyte, however, there is no unique salt variable and there are also multiple binary ambipolar diffusion coefficients.
If $N$ ionic species are present, the density perturbations contain one charge variable and $N-1$ independent electroneutral salt variables.
These neutral variables correspond to changes in composition at fixed charge density.
The impedance of a mixed electrolyte therefore need not be equivalent to that of a binary electrolyte with an effective diffusion coefficient or an effective conductivity.

For the limiting case of equal diffusivities, \(\epsilon_i=1\), the charge--salt formulation simplifies considerably.
The matrix $\boldsymbol{\Lambda}$ [\cref{eq:lambda}] governing the spatial modes has one charge eigenvalue $\lambda = 1+\tilde{p}$, and $N-1$ degenerate salt eigenvalues with $\lambda=\tilde{p}$.
The degeneracy of the salt modes reflects the fact that, when all ions diffuse at the same rate, all neutral salt combinations are dynamically equivalent.
For this case, the charge variable evolves independently of the salt variables, and the salt fluxes do not couple back to the electric field.
Consequently, the potential and impedance are identical to those of a binary electrolyte with equal cation and anion diffusivities\cite{Macdonald1953,usler_arxiv_2026}.
The additional salt degrees of freedom are present in the mathematical description, but they remain invisible in the impedance response.

\begin{figure}
    \centering
	\includegraphics[width =1.0\linewidth]{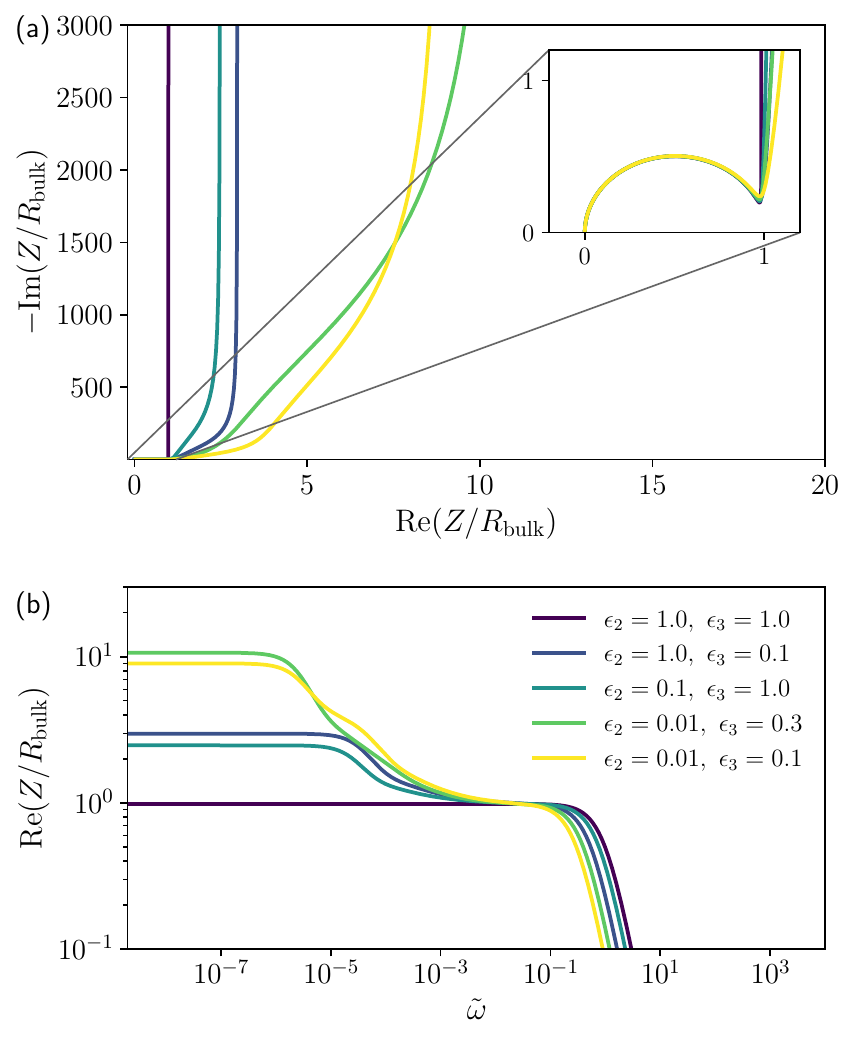}
    \caption{Impedance obtained from \cref{eq:sol_imp} for a ternary mixture with monovalent ions and $X_1=1/4$, $X_2=1/4$, and $X_3=1/2$ (a) in the Nyquist representation, (b) real part of impedance normalized by the bulk resistance $\mathrm{Re}(Z/R_{\mathrm{bulk}})$ as a function of dimensionless frequency $\tilde{\omega}$ for various diffusivities at $\tilde{L}=100$. The various colors correspond to different $(\epsilon_2,\epsilon_3)$ pairs.}\label{fig:noequal}
\end{figure}

The corresponding impedance has the well-known structure~\cite{Macdonald1953}, cf.~\cref{fig:noequal}(a), dark purple curve for a ternary electrolyte.  
In the Nyquist representation, a semicircular arc appears at high frequencies for $\tilde{L}=100$, while the low-frequency response approaches a vertical line at $\mathrm{Re}(Z/R_{\mathrm{bulk}})\approx 1$.
The latter is the capacitive regime, where the response is dominated by charge accumulation at the blocking electrodes.
In the frequency representation on \cref{fig:noequal}(b) with $\tilde{p}=i\tilde{\omega}$, \(\mathrm{Re}(Z/R_{\mathrm{bulk}})\) decreases from its low-frequency plateau towards the high-frequency resistance, consistent with the
transition from electrode-polarization dominated behavior to bulk transport.

The situation changes once one of the ions has a diffusivity different from the other two ions.
In ternary electrolytes, this case is relevant for mixtures such as NaCl--KCl, since the diffusion coefficients of $\mathrm{K}^+$ and $\mathrm{Cl}^-$ are relatively close, whereas $\mathrm{Na}^+$ diffuses more slowly.

I consider two representative cases $(\epsilon_2,\epsilon_3)=(1.0,0.1)\text{ and }(0.1,1.0)$.
In these cases, the salt degeneracy disappears.
One should no longer interpret the two salt variables as two identical, dynamically equivalent neutral modes.
Instead, the unequal diffusivities introduce a coupling between charge relaxation and one salt-relaxation mode.
The eigenvalue of the second salt-relaxation is again $\lambda=\tilde{p}$.
This can be also seen from the PNP equation, as the second salt decouples itself from the other two bases.
For a general multi-component electrolyte, one finds the same: the charge relaxation is coupled to one salt-relaxation mode, and the rest of the $N-2$ salt variables remain decoupled.
These systems effectively behave as a binary electrolyte with unequal diffusivities.

The charge-salt coupling is discernible in the impedance [\cref{fig:noequal}(a), blue and teal curves].
As the impedance has been normalized by the bulk resistance, the semicircular arc at high frequencies does not change with the variation of $\epsilon_2$ or $\epsilon_3$.
The low-frequency behavior remains predominantly capacitive.
This is expected for blocking electrodes: at sufficiently low frequency, ions have enough time to redistribute near the interfaces, and the response is controlled by the charging of the diffuse layers rather than by ordinary bulk conduction.
The vertical Nyquist branch is therefore preserved when the diffusivities are unequal, and the limiting value at $\tilde{\omega}\to0$ is different, because the two species have different bulk concentrations.
At intermediate frequencies, one sees Warburg-like features: 
there is a slanted region between the resistive arc and the capacitive vertical, with a varying slope and region width, which resembles a binary electrolyte with unequal diffusivities.
A more detailed discussion of binary systems with unequal diffusivities can be found in Ref.~\cite{usler_arxiv_2026}.

In case of a ternary electrolyte, when three diffusivities are different, the coupling between charge and salt degrees of freedom, as well as salt--salt coupling becomes stronger.
This can be observed from \cref{eq:final}, as for unequal diffusivities, neither of the terms drop.
Two exemplary cases are shown on \cref{fig:noequal}, $(\epsilon_2,\epsilon_3)=(0.01,0.3)\text{ and }(0.01,0.1)$ (green and yellow curves).
The resistive arc as well as the capacitive high freuency limit is unchanged.
The broadening of the spectra in the Nyquist representation can be observed because the diffusion coefficients differ more in magnitude.
For these two cases, the Warburg-like region has two distinguishable slopes [especially for $(0.01,0.1)$].
This suggests that the response is governed by two distinct diffusive relaxation modes, which can be interpreted as arising from two ambipolar diffusion coefficients.
Additionally, the slanted regions are not easily distinguishable when the diffusion coefficients are close to each other in value; the different slopes noticeably appear when the diffusion coefficients differ by roughly an order of magnitude.

Overall, the system with equal diffusion coefficients behaves effectively as a binary
electrolyte because the charge mode decouples from the neutral salt modes.
Unequal diffusivities break this simplification.
The impedance then contains information not only about charge relaxation, but also about how salt redistribution is coupled to the electric field.
This makes multi-component electrolytes qualitatively different from a single binary electrolyte, even when the ions are monovalent and the system is electroneutral in the bulk.

\section{Outlook and Conclusion}\label{sec:concl}
I derived expressions for the linearized impedance response of a multi-component electrolyte in the PNP framework and analyzed an exemplary ternary electrolyte for various diffusion coefficient ratios.
I found that the impedance response becomes mathematically more complex when the diffusivities of the ions are unequal due to the emerging charge--salt and possibly, salt--salt coupling.
When components have the same diffusion coefficient, the corresponding salt modes do not affect in the impedance response.
In the Nyquist representation, the impedance exhibits capacitive behavior at low frequencies, resistive behavior at high frequencies, and, at intermediate freuqencies, depending on the diffusion coefficient ratios, one finds a Warburg-like response, related to the ambipolar diffusion coefficient(s) of the system.

The multi-component model can be extended with Faradaic currents, or with ionic correlations via the Bazant-Storey-Kornyshev (BSK) equation~\cite{bazant_prl_2011} and implementing a similar approach of solving the BSK-NP equations as in Ref.~\cite{fertig_pre_2025}, although the mathematical complexity of such systems might make analytical insight very difficult, or even impossible to obtain.

\section*{Conflicts of Interest}
There are no conflicts of interest to declare.

\section*{Data availability}
All routines used in the creation of the figure is available at link: xx.

\section*{Acknowledgement}
This work was supported by a FRIPRO grant from The Research Council of Norway (Project No. 345079) and the EU's Horizon Europe research and innovation programme under GA No. 101137725 (BatCAT). I furthermore thank Mathijs Janssen and Adrian Usler for the fruitful conversations.
\appendix
\renewcommand\thefigure{\thesection\arabic{figure}} \setcounter{figure}{0}

\section{Derivation of \cref{eq:matrix_eta} for $N=2,3$}\label{appendix:eta}
For a binary electrolyte, there is only one salt variable, for which the boundary condition $\partial_{\tilde{x}}\hat{\tilde{s}}=0$ gives
\begin{align}\label{eq:bcsalt}
A_1k_1\cosh(k_1\tilde{L})\alpha_{12}+A_2k_2\cosh(k_2\tilde{L})\alpha_{22}=0
\end{align}
and therefore
\begin{align}
    A_2 = -A_1\dfrac{k_1\cosh(k_1\tilde{L})}{k_2\cosh(k_2\tilde{L})}\dfrac{\alpha_{12}}{\alpha_{22}}.
\end{align}
As $\boldsymbol{\alpha}_s=(\alpha_{12},\alpha_{22})$, $\boldsymbol{\alpha}_s^{(1)}=\alpha_{22}$ and $\boldsymbol{\alpha}_s^{(2)}=\alpha_{12}$, and their determinants is identical to their only element.

For a ternary system, we have two salt variables with boundary conditions $\partial_{\tilde{x}}\hat{\tilde{s}}_i=0$ with $i=1,2,3$, yielding
\begin{subequations}\label{eq:saltsalt}
\begin{align}
    \sum\limits_iA_ik_i\cosh(k_i\tilde{L})\alpha_{i2}&=0\\
    \sum\limits_iA_ik_i\cosh(k_i\tilde{L})\alpha_{i3}&=0
\end{align}    
\end{subequations}
With a little algebra, $A_3$ is then expressed as
\begin{align}\label{eq:A3}
    A_3 = -A_1\dfrac{k_1\cosh(k_1\tilde{L})}{k_3\cosh(k_3\tilde{L})}\dfrac{\alpha_{13}\alpha_{22}-\alpha_{12}\alpha_{23}}{\alpha_{22}\alpha_{33}-\alpha_{23}\alpha_{32}}
\end{align}
which gives
\begin{align}\label{eq:A2}
    A_2 = -A_1\dfrac{k_1\cosh(k_1\tilde{L})}{k_2\cosh(k_2\tilde{L})}\dfrac{\alpha_{12}\alpha_{33}-\alpha_{13}\alpha_{32}}{\alpha_{22}\alpha_{33}-\alpha_{23}\alpha_{32}}.
\end{align}
$\boldsymbol{\alpha}_s$ is a $2\times 3$ matrix, therefore it is easy to confirm that \cref{eq:A3,eq:A2} have the determinants of the sub-matrices of $\boldsymbol{\alpha}_s$ in the numerator and denominator.
The matrix $\boldsymbol{\eta}$ then can be constructed.
For a quaternary or a general multi-component system, evaluating $A_i$ with $i=2,\dots,N$ ($N\ge 4$) are done analogously, with using the steps shown above, although with increasing mathematical complexity for the calculation of the determinant. 

\section{Matrix equation for arbitrary valencies and composition in a ternary system}\label{appendix:general}
Following from \cref{eq:nondimpoiss}, one can construct the charge--salt differential equations for arbitrary valencies and compositions with the $q=z_1\rho_1+z_2\rho_2+z_3\rho_3$, $s_1=-z_3X_3\rho_1+z_1X_1\rho_3$, and $s_2=-z_3X_3\rho_2+z_2X_2\rho_3$, in dimensionless form as
\begin{subequations}
    \begin{align}
    \partial_{\tilde{t}} \tilde{q} &=  C_{11}\partial_{\tilde{x}}^2\tilde{q}+C_{12}\partial_{\tilde{x}}^2\tilde{s}_1+C_{13}\partial_{\tilde{x}}^2\tilde{s}_2 +C_{11}\partial^2_{\tilde{x}}\tilde{\psi}\\
    \partial_{\tilde{t}} \tilde{s}_1 &= C_{21}\partial_x^2\tilde{q}+C_{22}\partial_{\tilde{x}}^2\tilde{s}_1+C_{23}\partial_{\tilde{x}}^2\tilde{s}_2 +C_{21}\partial^2_{\tilde{x}}\tilde{\psi}\\
    \partial_{\tilde{t}} \tilde{s}_2 &= C_{31}\partial_{\tilde{x}}^2\tilde{q}+C_{32}\partial_{\tilde{x}}^2\tilde{s}_1+C_{33}\partial_{\tilde{x}}^2\tilde{s}_2  +C_{31}\partial^2_{\tilde{x}}\tilde{\psi},
    \end{align}
\end{subequations}
with
\begin{subequations}
    \begin{align}
    C_{11}&=\dfrac{X_1z_1^2+X_2z_2^2\epsilon_2+X_3z_3^2\epsilon_3}{S}\\
    C_{12}&=\dfrac{z_1[X_2z_2^2(\epsilon_2-1)+X_3z_3^2(\epsilon_3-1)]}{X_3z_3S}\\
    C_{13}&=\dfrac{z_2[X_1z_1^2(1-\epsilon_2)+X_3z_3^2(\epsilon_3-\epsilon_2)]}{X_3z_3S}\\
    C_{21}&=\dfrac{X_1X_3z_1z_3(\epsilon_3-1)}{S}\\
    C_{22}&=\dfrac{(X_2z_2^2+X_3z_3^2)+X_1z_1^2\epsilon_3}{S}\\
    C_{23}&=\dfrac{X_1z_1z_2(\epsilon_3-1)}{S}\\
    C_{31}&=\dfrac{X_2X_3z_2z_3(\epsilon_3-\epsilon_2)}{S}\\
    C_{32}&=\dfrac{X_2z_1z_2(\epsilon_3-\epsilon_2)}{S}\\
    C_{33}&=\dfrac{(X_1z_1^2+X_3z_3^2)\epsilon_2+X_2z_2^2\epsilon_3}{S}.
    \end{align}
\end{subequations}
where $S=\sum_iz_i^2X_i$.
From hereon, after Laplace transformation, following the same mathematical steps described in \cref{sec:tern}, one finds the eigenvalues of matrix $\mathbf{M}$.
The solution for the Laplace-transformed potential and impedance is then identical, with $R_{\mathrm{bulk}}=\tilde{L}(z_1^2X_1+z_2^2X_2+z_3^2X_3)/(z_1^2X_1+\epsilon_2 z_2^2X_2+\epsilon_3 z_3^2X_3)$.
\bibliography{bibliography}
\bibliographystyle{unsrt}
\end{document}